\documentclass{emulateapj}

\newcommand{\lsun}{\mbox{L$_\odot$}}
\newcommand{\msun}{\mbox{M$_\odot$}}
\newcommand{\rsun}{\mbox{R$_\odot$}}

\newcommand{\tbol}{\mbox{$T_{bol}$}} 

\newcommand{\mdisk}{\mbox{$M_{disk}$}} 
\newcommand{\menv}{\mbox{$M_{env}$}} %
\newcommand{\rdisk}{\mbox{$R_{disk}$}} %
\newcommand{\rcent}{\mbox{$R_{cent}$}} %
\newcommand{\rout}{\mbox{$R_{out}$}} %
\newcommand{\ang}{\mbox{Ang}} 
\newcommand{\incl}{\mbox{Incl}} 

\shorttitle{The Resolved Massive Disk in FIRS 1}
\shortauthors{Enoch et al.}

\begin{document}

\title{Disk and Envelope Structure in Class 0 Protostars: I. The Resolved Massive Disk in Serpens FIRS~1}

\author{Melissa L. Enoch (1,2), Stuartt Corder (3,4), Michael M. Dunham (5), and Gaspard Duch\^{e}ne (1,6)}

\affil{
(1) Department of Astronomy, University of California at Berkeley, 601 Campbell Hall, Berkeley, CA, 94720 \\
(2) \textit{Spitzer} Fellow; menoch@berkeley.edu \\
(3) NRAO/ALMA-JAO, Av. Apoquindo 3650, Piso 18, Las Condes, Santiago, Chile \\
(4) Jansky Fellow, NRAO \\
(5) Department of Astronomy, The University of Texas at Austin, 1 University Station, C1400, Austin, TX 78712 \\
(6) Universit\'e Joseph Fourier - Grenoble 1 / CNRS,  Laboratoire d'Astrophysique de Grenoble (LAOG) UMR 5571, BP 53, 38041 Grenoble Cedex 09, France
}

\begin{abstract}

We present the first results of a program to characterize the disk and envelope structure of typical Class~0 protostars in nearby low-mass star forming regions.
We use \textit{Spitzer} IRS mid-infrared spectra, high resolution CARMA 230~GHz continuum imaging, and 2-D radiative transfer models to constrain the envelope structure, as well as the size and mass of the circum-protostellar disk in Serpens FIRS~1.
The primary envelope parameters (centrifugal radius, outer radius, outflow opening angle, and inclination) are well constrained by the spectral energy distribution (SED), including \textit{Spitzer} IRAC and MIPS photometry, IRS spectra, and 1.1~mm Bolocam photometry.  These together with the excellent $uv$-coverage ($4.5-500$~k$\lambda$) of multiple antenna configurations with CARMA allow for a robust separation of the envelope and a resolved disk.
The SED of Serpens FIRS~1 is best fit by an envelope with the density profile of a rotating, collapsing spheroid with an inner (centrifugal) radius of approximately 600~AU, and the millimeter data by a large resolved disk with $\mdisk\sim1.0~ \msun$ and $\rdisk\sim 300$~AU.
These results suggest that large, massive disks can be present early in the main accretion phase.  Results for the larger, unbiased sample of Class~0 sources in the Perseus, Serpens, and Ophiuchus molecular clouds are needed to determine if relatively massive disks are typical in the Class~0 stage.

\end{abstract}

\section{Introduction}

Protostars build up their mass by accreting material from a dense protostellar envelope, presumably via a rotationally supported circum-protostellar accretion disk.
Disk formation is a natural result of collapse in a rotating core, but it is not known how soon after protostellar formation the disk appears, or how massive it is at early times.
Theory suggests that centrifugally supported disks should start out small (radius less than $10$~AU), and thus low mass, and grow with time \citep*{tsc84}. Unstable or magnetically supported disks, however, could be much larger (radii up to 1000~AU in the magnetically supported case; \citealt{gs93}), and thus more massive at early times.

The remnants of these protostellar accretion disks are easily observed in more evolved phases (e.g. TTauri stars), but given that the majority of mass is accreted during earlier embedded phases, understanding disks at early times is critical.  Directly observing disks during this main accretion phase is quite difficult, however, as they are hidden within dense, extincting protostellar envelopes.
The structure of the envelope at small radii is another important characteristic of main accretion phase protostars that is similarly difficult to directly observe.
Disk growth or the presence of a binary companion may clear out the inner region of the envelope early on, as inferred for the binary Class~0 source IRAS 16293-2422 by \citep{jorg05}.  

There has been a recent push toward detecting disks in more embedded objects, with many now known and roughly characterized in Class I protostars \citep[e.g.][]{loon00,jorg05b,eisner05,aw07}, and a few detected in the earlier Class 0 stage \citep[e.g.][]{chan95,harv03,brown00,loon00}.
\footnote{We use definitions of Class 0, Class I, and Class II \citep{andre94} based on the bolometric temperature \citep{ml93,chen95}: $\tbol \le 70$~K (Class~0); 70 K$< \tbol \le 650$~K (Class~I); 650 K$<\tbol \le 2800$ K (Class~II).  We further assume that classes correspond to an evolutionary sequence \citep[e.g.][]{rob06}: in Class~0 the protostar has accreted less than half its final mass ($M_{*} < \menv$), in Class~I $M_* > \menv$, and in Class~II the envelope has dispersed, leaving only a circum-stellar disk.}
Most previous detailed studies have been limited to the most well-known or brightest Class 0 sources, however, due to instrumental limitations and a lack of large unbiased target samples.  The ongoing Submillimeter Array survey of low-mass protostars \citep{jorg07,jorg09} is a notable exception.

With recent large surveys of nearby molecular clouds at mid-infrared and (sub)millimeter wavelengths it is now possible to define complete samples of Class 0 protostars based on luminosity or envelope mass limits \citep[e.g.][]{hatch07,jorg08,dun08,enoch09,evans09}.

\begin{figure*}[!ht]
\vspace{-0.2in}
\begin{center}
\includegraphics[width=4.8in]{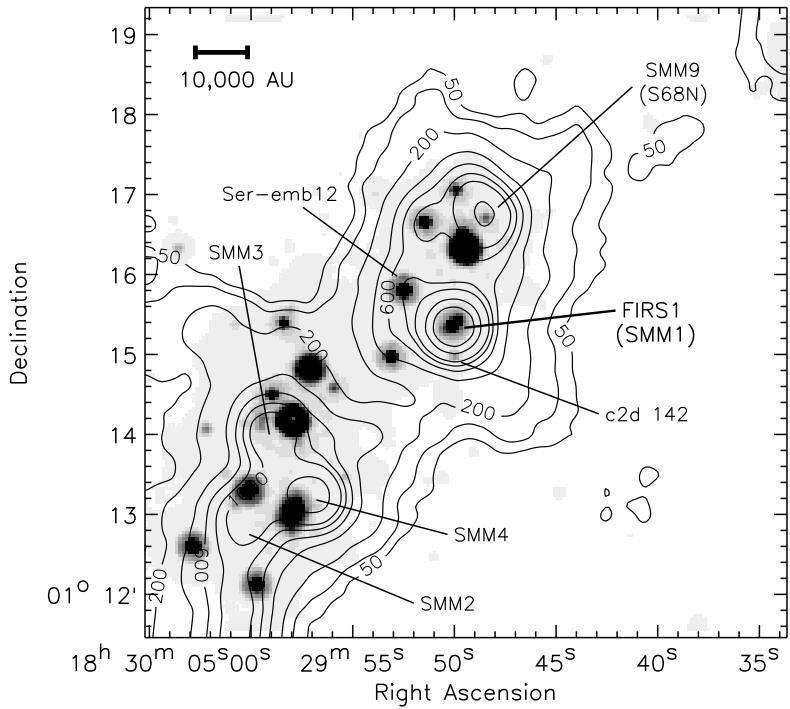}
\vspace{-0.2in}
\caption
{\textit{Spitzer} $24~\micron$ image of the immediate environment of Serpens FIRS~1, in the Serpens main core.  Bolocam 1.1~mm continuum contours are overlaid. 
Submillimeter source designations for the brightest \citet{casali93} sources are indicated.  The nearest embedded protostar to FIRS~1 is approximately $45\arcsec$ or $11000$~AU away (Ser-emb 12; \citealt{enoch09}), and the nearest YSO is $25\arcsec$ or 6000~AU away (c2d 142; \citealt{harv07}).
\label{genfig}}
\end{center}
\end{figure*}

We have recently begun a campaign to characterize disk properties in a large, uniform sample of Class 0 protostars in nearby low mass star-forming regions (M. Enoch et al. 2009, in preparation).
Our study is based on the complete (to envelope masses $\gtrsim0.1~\msun$) sample of 39 Class 0 protostars in the Serpens, Perseus, and Ophiuchus molecular clouds, identified by \citet{enoch09} by comparing large-scale \textit{Spitzer} IRAC and MIPS and Bolocam 1.1~mm continuum surveys of the three clouds.

Combining Spitzer IRS mid-infrared (MIR) spectra and high resolution CARMA 230~GHz continuum imaging with radiative transfer modeling of this sample will help to address several fundamental questions about the structure and evolution of the youngest protostars: 1) How soon after the initial collapse of the parent core does a circum-protostellar disk form? 2) What fraction of the total circum-protostellar mass resides in the disk, and does this fraction vary with time?  3) Are large ``holes'' in the inner envelope, such as that found for IRAS~16293 by \citet{jorg05}, common at early times?

MIR spectra and millimeter maps provide complementary approaches to these questions. The amount of flux escaping at $\lambda \lesssim 50~\micron$ from deeply embedded sources is very sensitive to the opacity close to the protostar, and thus the envelope structure \citep[e.g.][]{jorg05}. While the MIR flux is insensitive to disk properties, high resolution millimeter continuum mapping can directly detect emission from dust grains in the disk.  Millimeter observations with excellent $uv$-coverage, combined with radiative transfer models, can separate the disk from the envelope and constrain the disk mass and size.

Our ultimate goal is to characterize the disk mass, size, and inner envelope structure of typical low-mass Class~0 protostars, and to quantify any trends with evolutionary indicators.
In this initial paper we present results for Serpens FIRS~1, a well known Class 0 source, which will serve as a test case for the full program.  

\section{Serpens FIRS 1}

FIRS~1 is located at $18^h29^m49.6^s +01^o15'21.9''$ (J2000)\footnote{\textit{Spitzer} position from \citet{harv07}} in the main core (Cluster A) of the Serpens Molecular Cloud.  We adopt a distance of $d=260\pm10$~pc \citep{stra96}, and any quoted literature values are scaled to this distance.  It is a well know Class~0 protostar \citep[e.g.][]{hb96} first noted in the far-infrared by \citet*{hwj84}, and also known by its sub-millimeter designation Serpens SMM~1 \citep{casali93}.
There is a narrow $\lambda=3.6$~cm bipolar radio jet at the position of FIRS~1 \citep*{rod89,curiel93}, indicating a powerful outflow that is also clearly seen in molecular lines \citep{davis99,curiel96}.

Figure~\ref{genfig} gives an overview of the FIRS~1 environment with \textit{Spitzer} $24~\micron$ and Bolocam 1.1~mm continuum maps of the Serpens main core.  
The nearest known YSO is approximately $25\arcsec$ away, or 6000~AU in projected distance \citep{harv07}, and the nearest embedded protostar known to have an envelope is $45\arcsec \approx 11000$~AU away \citep{enoch09}.

\begin{figure*}[!ht]
\begin{center}
\includegraphics[width=5.4in]{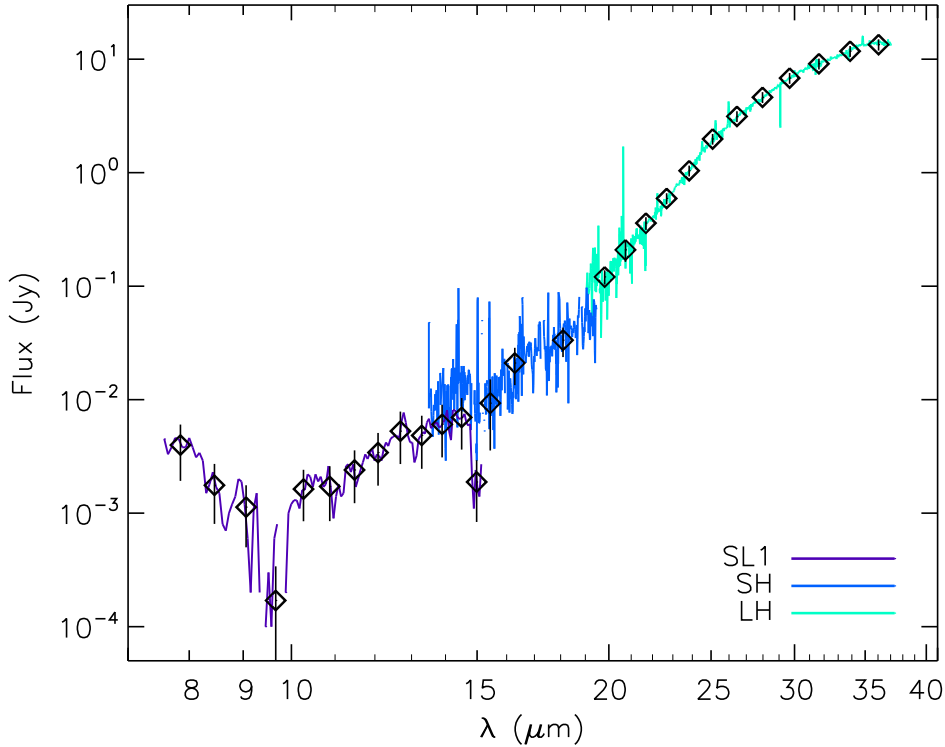}
\caption
{\textit{Spitzer} IRS spectrum of Serpens FIRS~1, using the Low Res $7.4-14.5~\micron$ (SL1), Hi Res $9.9-19.6~\micron$ (SH), and Hi Res $18.7-37.2~\micron$ (LH) modules.  The low signal-to-noise SH data at $\lambda<13.5~\micron$ is not plotted.  Binned data ($\Delta\lambda\sim 1~\micron$) are over-plotted as diamonds; error bars represent the error in the mean for each bin.  Binned fluxes are used in the model fitting and given in Table~\ref{irstab}.
\label{irsfig}}
\end{center}
\end{figure*}

FIRS~1 is referred to as Ser-emb~6 in \citet{enoch09}, and is
associated with the 1.1~mm Bolocam core Ser-Bolo~23 \citep{enoch07}.
Based on 2MASS, \textit{Spitzer}, and Bolocam data the bolometric
luminosity is $11.0~ \lsun$,\footnote{Note that lower resolution HIRES IRAS fluxes
  \citep{hb96} and ISO-LWS spectra \citep{larsson00} yield higher
  bolometric luminosities of $62~\lsun$ and $95~\lsun$, respectively, but these may be confused with nearby embedded protostars.}
the bolometric temperature is
$56$~K, confirming the Class~0 designation, and the total envelope
mass is $8.0 \msun$ \citep{enoch09}.

Previous high resolution millimeter observations have placed limits on the mass of a compact disk in FIRS~1. \citet*{hoger99} used observation from the Owens Valley Radio Observatory (OVRO) millimeter interferometer to estimate a total mass (disk plus envelope) within 100~AU of $0.7~\msun$.  \citet{brown00} place a lower limit on the disk mass of $\sim 0.1~\msun$ with sub-millimeter observations ($\nu\sim350$~GHz) from the James Clerk Maxwell Telescope-Caltech Submillimeter Observatory (JCMT-CSO) single baseline interferometer.

\section{Observations}\label{obssec}

\subsection{Spitzer IRS spectrum}\label{irsobs}

Mid-infrared spectra were obtained with the Infrared Spectrograph (IRS) on the \textit{Spitzer Space Telescope} \citep{werner04,houck04} during 2007 October 5 with the  Low Res $7.4-14.5~\micron$ (SL1; $R\sim100$),  Hi Res $9.9-19.6~ \micron$ (SH; $R\sim600$), and Hi Res $18.7-37.2~\micron$ (LH; $R\sim600$) modules.  
Integration times were 117 sec in SL1, 189 sec in SH, and 59 sec in LH.
Off-source or background spectra with the same integration times were also obtained for SH and LH.

Spectra were extracted from the SSC pipeline version S16.1.0 BCD images using the reduction pipeline \citep{lahuis06} developed for the ``From Molecular Cores to Planet-forming Disks'' \textit{Spitzer} Legacy Program (``Cores to Disks'' or c2d; \citealt{evans03}).
We use the optimal PSF extraction method of the pipeline, which is based on fitting the analytical cross dispersion point spread function, plus extended emission, interpolating over bad pixels.  The 1-D spectra are flux calibrated using a spectral response function derived from a suite of calibrator stars, corrected for instrumental fringe residuals, and an empirical order matching algorithm is applied.
PSF extraction was completed for both FIRS~1 and the background field; the final spectrum is the difference between them.  

The resulting spectrum from $7.4$ to $37.2~\micron$ is shown in Figure~\ref{irsfig}.  Each module is reduced separately, so the degree of agreement gives some idea of the reliability of calibration.  
Both the full reduced spectra (solid lines), and the average flux in wavelength bins of $\Delta \lambda \sim 0.75~\micron$, $1.5~ \micron$, and $1~\micron$ for SL1, SH, and LH, respectively, are shown.  The SH spectrum has the lowest signal to noise, so it is binned on the coarsest grid.  
Only data with signal-to-noise greater than one (SL1, SH) or three (LH) are included in the binned points.  Binned fluxes are listed in Table~\ref{irstab}.

In Figure~\ref{irsfig} the silicate absorption band at $9.7~\micron$ is clearly visible, and a hint of the CO$_2$ ice band at $15~\micron$ is also visible.  A number of finer features in the LH spectrum are most likely real, but will not be discussed here.

\begin{figure*}[!ht]
\vspace{-0.3in}
\includegraphics[width=7.2in]{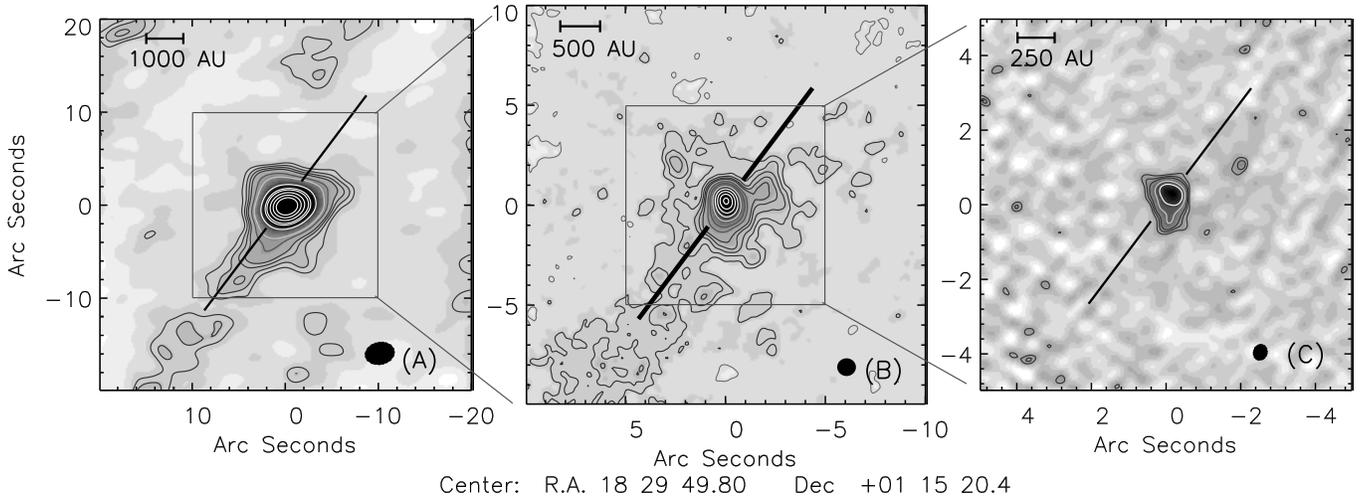}
\vspace{-0.3in}
\caption
{CARMA 230 GHz maps of Serpens FIRS~1 for short baseline data only (D,E configurations; panel A), all data (panel B), and long baseline data only (B,C configurations; panel C).  Contours in panel B are (2,4...10,15,20,30...70) times the $1\sigma$ rms of 6.7 mJy~beam$^{-1}$, for a synthesized beam of $0.94\arcsec \times 0.89\arcsec$ (shown, lower right).  Contours in panels A and C are similar but start at $4\sigma$ and $6\sigma$, respectively, and panel A has additional contours at ($90,110,130~\sigma$).  Note the change in scale in each panel.  
The direction of the 3.6~cm jet \citep{rod89,curiel93} is shown for reference.
\label{carmafig}}
\vspace{0.1in}
\end{figure*}

\subsection{CARMA 230 GHz map}\label{carmaobs}

Continuum observations at $\nu=230$~GHz ($\lambda=1.3$~mm) were completed with the Combined Array for Research in Millimeter-wave Astronomy (CARMA), a 15 element interferometer consisting of nine 6.1~m and six 10.4~m antennas.
Data were obtained in the B-array ($100-1000$~m baselines), C-array ($30-350$~m), D-array ($11-150$~m), and E-array ($8-66$~m) configurations between 2007 October 24 and 2008 December 31.  These data were combined to provide $uv$-coverage from $4.5$ to $500~ k\lambda$.  
Small 7-pointing mosaics were made in the compact configurations (D and E) in order to mitigate spatial filtering by the interferometer, and to more fully map the spatially extended protostellar envelope.

All three correlator bands were configured for continuum, 468 MHz bandwidth, observations.  
A bright quasar (1751+096) was observed approximately every 15 minutes to be used for complex gain calibration.  Absolute flux calibration was accomplished using 5 minute observations of Uranus, Neptune, or MWC~349.  The overall calibration uncertainty is approximately 20\%, from the reproducibility of the phase calibrator flux on nearby days. 
A passband calibrator, typically 3C454.3, was observed for 15 minutes during each set of observations, and radio pointing was performed every two hours.  
Observations in the most extended B configuration utilized the Paired Antenna Calibration System (PACS) to correct for phase variations on minute timescales (see L. P\'{e}rez et al. 2009, in preparation).

Calibration and imaging were accomplished with the MIRIAD data reduction package \citep*{sault95}.  The resulting 230~GHz map of FIRS~1 is shown in Figure~\ref{carmafig}, with maps made at three resolutions:  short baseline data only (D and E configurations), all data, and long baseline data only (B and C configurations).  The direction of the 3.6~cm jet \citep{rod89,curiel93} is shown for reference.

The map including all data was inverted with natural weighting, cleaned with a Steer CLEAN algorithm \citep*{steer84}, and restored with an $0.94\arcsec \times 0.89\arcsec$ beam.  The rms noise level in the central region is 6.7~mJy~beam$^{-1}$, the peak and total flux from a Gaussian fit are $0.42$ Jy beam$^{-1}$ and 1.38 Jy (PA$=-4\deg$), and the deconvolved FWHM size is $1.6\arcsec \times1.2\arcsec$.  The synthesized beam corresponds to approximately 240~AU, while the longest baselines provide a resolution better than 100~AU ($0.46\arcsec \times0.40\arcsec$).  A gaussian fit to the long baseline data yields PA$=25\deg$, approximately $75\deg$ from the 3.6~cm jet axis (PA$=-50\deg$).

\begin{figure*}[!ht]
\begin{center}
\includegraphics[width=5.5in]{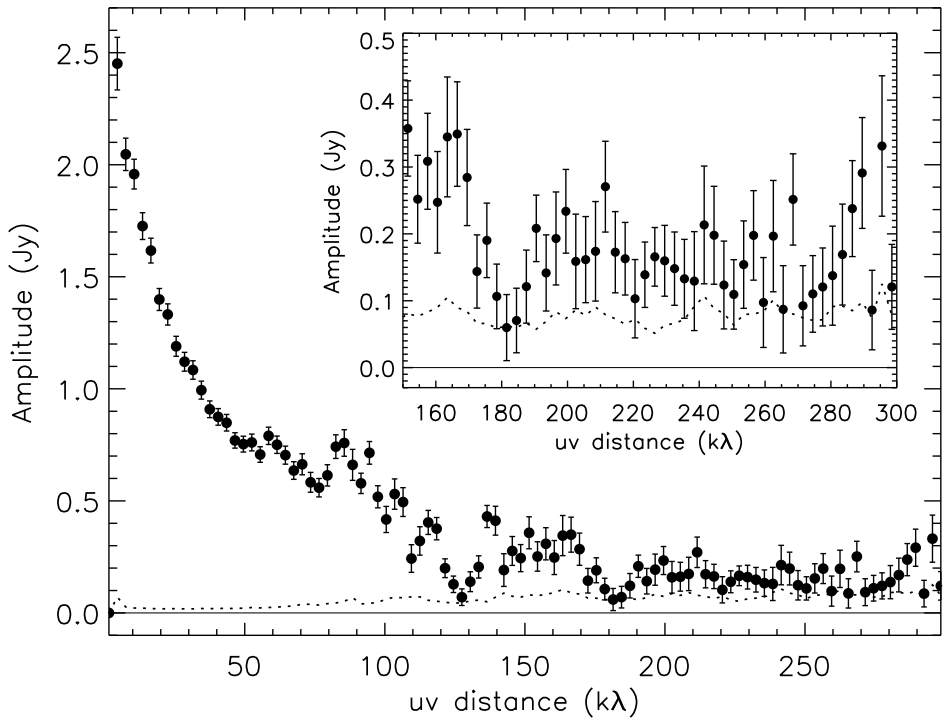}
\caption
{CARMA 230~GHz visibility amplitude versus $uv$-distance for Serpens FIRS~1.  Observations in the B, C, D, and E CARMA antenna configurations provide $uv$-coverage from approximately $4.5~k\lambda$ to $500~ k\lambda$.  The expected value in the case of zero signal, or amplitude bias, is indicated by a dotted line and is typically small (less than 0.1~Jy).
\label{visfig}}
\end{center}
\end{figure*}

The extended, complex nature of the source is apparent, thanks to the excellent $uv$-coverage achieved with multiple configurations.  Although it is difficult to see the more extended envelope even in the short baseline map, it is clearly visible as an amplitude peak at $uv$-distances $<20~k\lambda$ in a plot of amplitude versus $uv$-distance (Figure~\ref{visfig}).  
Note that the interferometer does filter out flux at $uv$-distances less than $4~ k\lambda$, corresponding to the separation of the closest antenna pairs.  Figure~\ref{visfig} shows that  most of the source flux is concentrated at low and intermediate $uv$-distances (extended structure), but the source is clearly detected at $uv$-distances greater than $200~k\lambda$, indicating an unresolved or marginally resolved compact ($<1''$) component. 
Values of the 230~GHz flux as a function of $uv$-distance are given in Table~\ref{vistab}.

\input{tab1}
\input{tab2}

\subsection{Spitzer, Bolocam, and SHARC II broadband data}\label{broadbandobs}

Broadband infrared data for FIRS~1 are taken from the ``Cores to Disks''  \textit{Spitzer} Legacy program \citep{evans03}, which imaged approximately 1 square degree in the cloud with IRAC and MIPS \citep{harv06,harv07}.  The same region was mapped at $\lambda=1.1$~mm with the Bolocam bolometer array \citep{glenn03} at the Caltech Submillimeter Observatory (CSO) \citep{enoch07}.
These data provide wavelength coverage from $\lambda=3.6$ to $1100~ \micron$ (IRAC 3.6, 4.5, 5.8, 8.0~$\micron$; MIPS 24, 70, 160~$\micron$; Bolocam 1100~$\micron$).  FIRS~1 is not detected in the 2MASS catalogs.

Broadband fluxes are used to determine the bolometric luminosity and temperature ($11.0~ \lsun$ and $56$~K), and are included in the model fits in Section~\ref{ressec}, below.  The total envelope mass ($8.0~ \msun$) is calculated from the total flux in a $40\arcsec$ aperture at $\lambda=1.1$~mm, assuming the envelope is optically thin at 1.1~mm, a dust opacity of $\kappa_{1mm} = 0.0114$~cm$^2$~g$^{-1}$ \citep{oh94}, and a dust temperature of $T_D=15$~K (see \citealt{enoch09} for more details).

We also include in the observed SED the $350~\micron$ continuum flux (M. Dunham et al. 2009, in preparation), obtained with SHARC-II \citep{dowell03} at the CSO.  The SHARC-II flux samples the peak of the SED, and helps constrain the long-wavelength side of the model SED.  
All fluxes used in the SED fit are given in Table~\ref{irstab}, including uncertainties, aperture diameters and instrument used for the observations.

\section{Radiative transfer model}\label{modsec}

To model the observed emission from FIRS~1, we use the two-dimensional Monte Carlo radiative transfer code RADMC of \citet{dd04}.  
RADMC performs both Monte Carlo radiative transfer 
to derive the temperature distribution from an input density distribution, and ray tracing to produce images and photometry in specified apertures.
We adopt a density profile very similar to that of \citet{crapsi08}, which includes three components: the envelope, the outflow cavity, and the disk.
For both the envelope and disk we use the dust opacities from Table 1, column 5 of
\citet{oh94} for dust grains with thin ice mantles, including scattering, interpolated onto the necessary wavelength grid.
We note that although this is a young source, there could be significant difference in the dust properties of the disk and envelope, as have been demonstrated in some Class~I sources \citep{wolf03}.

The envelope density profile is that of a rotating, collapsing sphere \citep{ulrich67}:
\begin{equation}
\rho_{env}(r,\theta) = \rho_0  \left(\frac{r}{R_c}\right)^{-1.5} \left(1+\frac{\mu}{\mu_0}\right)^{-0.5} \left( \frac{\mu}{\mu_0}+2\mu_0^2 \frac{R_c}{r}\right)^{-1},
\end{equation}
where $\mu=\cos\theta$, $R_c=\rcent$ is the centrifugal radius, and $\rho_0$ is the density at $(r,\theta)=(R_c,0)$, which is set by the total envelope mass $\menv=8~\msun$.  A gas-to-dust mass ratio of $100$ and mean molecular weight of 2.33 are included.  
Here $\mu_0=\cos\theta_0$ is the solution of the parabolic motion of an infalling particle, given by:
\begin{equation}
\left(\frac{r}{\rcent}\right) \left(\frac{\mu_0 - \mu}{\mu_0 \sin^2\mu_0}\right) = 1
\end{equation}

The model has no time dependence; \rcent\ is not a true centrifugal radius defined by the conservation of angular momentum during collapse, but simply a set radius where the density peaks, and inward of which the density drops to very low values. 
The envelope outer radius \rout\ is the maximum value of the radial grid, so the envelope density is zero for $r>\rout$.
The outflow cavity is defined by setting the density to zero in the region where $\cos\theta_0>\cos(\ang/2)$.  This results in a funnel-shaped cavity, which is conical only at large scales where $\ang$ is the full opening angle.

The disk density is given by a power law dependence in radius and a Gaussian dependence in height:
\begin{equation}
\rho_{disk}(r,\theta) = \frac{\Sigma_0}{\sqrt{2\pi} H(r)}  \left(\frac{r}{\rdisk}\right)^{p1} \exp \left[-\frac{1}{2}\left(\frac{r \mu}{H(r)}\right)^2\right],
\end{equation}
where $\Sigma_0$ is set by the input disk mass \mdisk.
At \rdisk\, $p1$ changes from $-1$ to $-12$, effectively setting the disk radius.  
The scale-height variation (flaring) is given by: $H(r)=r (H_0/\rdisk)(r/\rdisk)^{p2}$. We set $p2=2/7$, corresponding to the self-irradiated passive disk of \citet{cg97}.
Given that the disk is in large part hidden by the envelope, a much simpler description would likely work just as well, but we choose to follow the setup of \citet{crapsi08} here.  Most of the disk parameters are held fixed for the main model grid, with only \mdisk\ and \rdisk\ varying.

Table~\ref{paramtab} lists the range of input values for the model input parameters, some of which are held fixed.
The internal luminosity is set by the bolometric luminosity; it is input to the model as the stellar luminosity, although most likely a majority of the luminosity is due to accretion.
We do not include an interstellar radiation field here, as the luminosity of any reasonable field is negligible compared to the internal luminosity of FIRS~1, and has only a negligible effect on the long-wavelength SED.

\begin{deluxetable*}{lccl}
\tablecolumns{4}
\tablewidth{0pc}
\tablecaption{\label{paramtab}Range of parameter values used in the radiative transfer model grid}
\tablehead{
\colhead{Parameter}  & \colhead{Fixed?} & \colhead{Range} & \colhead{Description}}
\startdata
\hline
\multicolumn{4}{c}{Protostar}  \\
\hline
$L_{star}$  & Y &  $11~ \lsun$  & Internal luminosity \\
$T_{star}$  & Y & $4000$~K  &  Protostar effective temperature \\
\hline
\multicolumn{4}{c}{Envelope and outflow}  \\
\hline
$\menv$  &  Y & $8.0~ \msun$   & Total mass of envelope \\
\rout\  & N &  $3000 - 12000$ AU   & Outer radius of envelope \\
\rcent\  & N &  $50 - 1000$ AU   & Centrifugal (inner) radius of envelope \\
\ang  & N &  $5 - 80$ deg  &  Outflow opening angle \\
\incl  & N &  $5 - 90$ deg  &  Inclination angle \\
\hline
\multicolumn{4}{c}{Disk}  \\
\hline
\mdisk\  & N &  $0.0 - 3.0~ \msun$   & Disk mass \\
\rdisk\  & N &  $50 - 1000$~AU   & Disk radius \\
$H_0$  & Y &  $0.2~ \rdisk$   & Disk vertical pressure scale height \\
$p1$  & Y &  $-1.0$  & Disk surface density radial power law ($r<\rdisk$) \\
$p2$  & Y &  $2/7$  & Power law for H(R) (disk flaring) 
\enddata
\tablecomments{The internal luminosity is set by the bolometric luminosity of the source, determined from the broadband SED, and the envelope mass is set by the 1.1~mm Bolocam single dish flux (see \citealt{enoch09}).  \ang\ is the full outflow opening angle.  \incl\ is the line of sight inclination angle of the disk: $0\deg$ is face-on, $90\deg$ is edge-on.  Stellar, envelope, and disk parameters are discussed in Section~\ref{modsec}.}
\end{deluxetable*}

Small single parameter grids were run to test that ``fixed'' parameters have no significant affect on the model SED or millimeter visibilities.  
The scale height ($H_0$) and flaring ($p1$) of the disk do affect the $3-20~\micron$ fluxes, although only for $\incl<10 \deg$, where there is too much NIR flux to match the data regardless of the value of these parameters.  A very puffy disk ($H_0\gtrsim0.5 \rdisk$) produces more emission in this range, while a very thin disk ($H_0\lesssim0.1 \rdisk$) produces less emission.  Similarly, a flared disk produces more MIR emission than one with no flaring.

\section{Results}\label{ressec}

To determine the best fit envelope parameters we run a grid of models varying \rout, \rcent, \ang, and \incl.  This results in a total of 588 envelope models, each observed at 13 inclination angles.  A nominal disk of $\mdisk=0.01~\msun$, $\rdisk = 150$~AU is used for all envelope models.  A separate grid varying \mdisk\ and \rdisk\ with fixed envelope parameters includes 140 models.  Our tests show that the disk has little effect on the SED for this source, making separate grids feasible. 

\begin{figure*}[!ht]
\includegraphics[angle=90,width=7.1in]{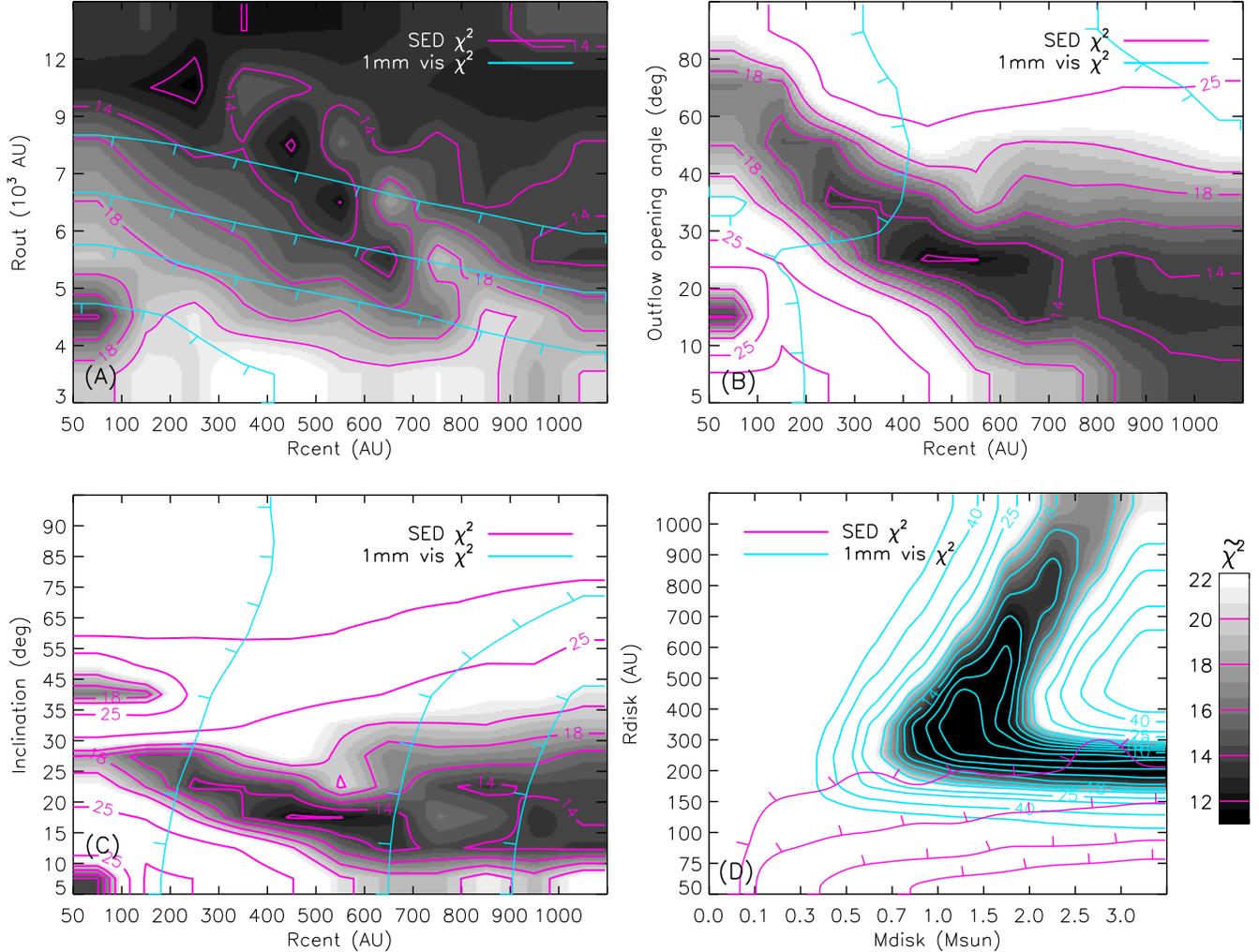}
\vspace{-0.15in}
\caption
{Reduced $\chi^2$ contours resulting from fitting the observed SED and CARMA 230~GHz visibilities (Tables~\ref{irstab} and \ref{vistab}) to the grid of envelope and disk models.  Models have been run for the full parameter ranges shown; the model grid resolution corresponds to the axis labels (e.g. \rcent\ values of $50, 100, 200, 300...1000$~AU), but the $\tilde{\chi}^2$ distribution has been smoothed for a better visual representation.  Contours from fits to the both the SED (magenta) and visibilities (cyan, tick marks indicate downhill direction) are shown, although envelope parameters (panels A-C) are primarily constrained by the SED, while disk parameters (panel D) are constrained by the millimeter visibilities. In panels (A)-(C) the $\tilde{\chi}^2$ distribution is collapsed along the parameters not plotted. The lowest contour plotted is $\tilde{\chi}^2=8$.
\label{chifig}}
\vspace{0.1in}
\end{figure*}

The model grids are compared to the SED and 230 GHz visibilities with a $\chi^2$ analysis using data from Tables~\ref{irstab} and \ref{vistab}.  
Contour plots of the resulting reduced $\chi^2$ ($\tilde{\chi}^2$)
are shown in Figure~\ref{chifig}, where contours for both the SED and 230~GHz visibilities are plotted.  The model and data visibilities are calculated by the same method, using vector averaging in radial annuli.

\begin{figure*}[!ht]
\hspace{-0.2in}
\includegraphics[width=7.3in]{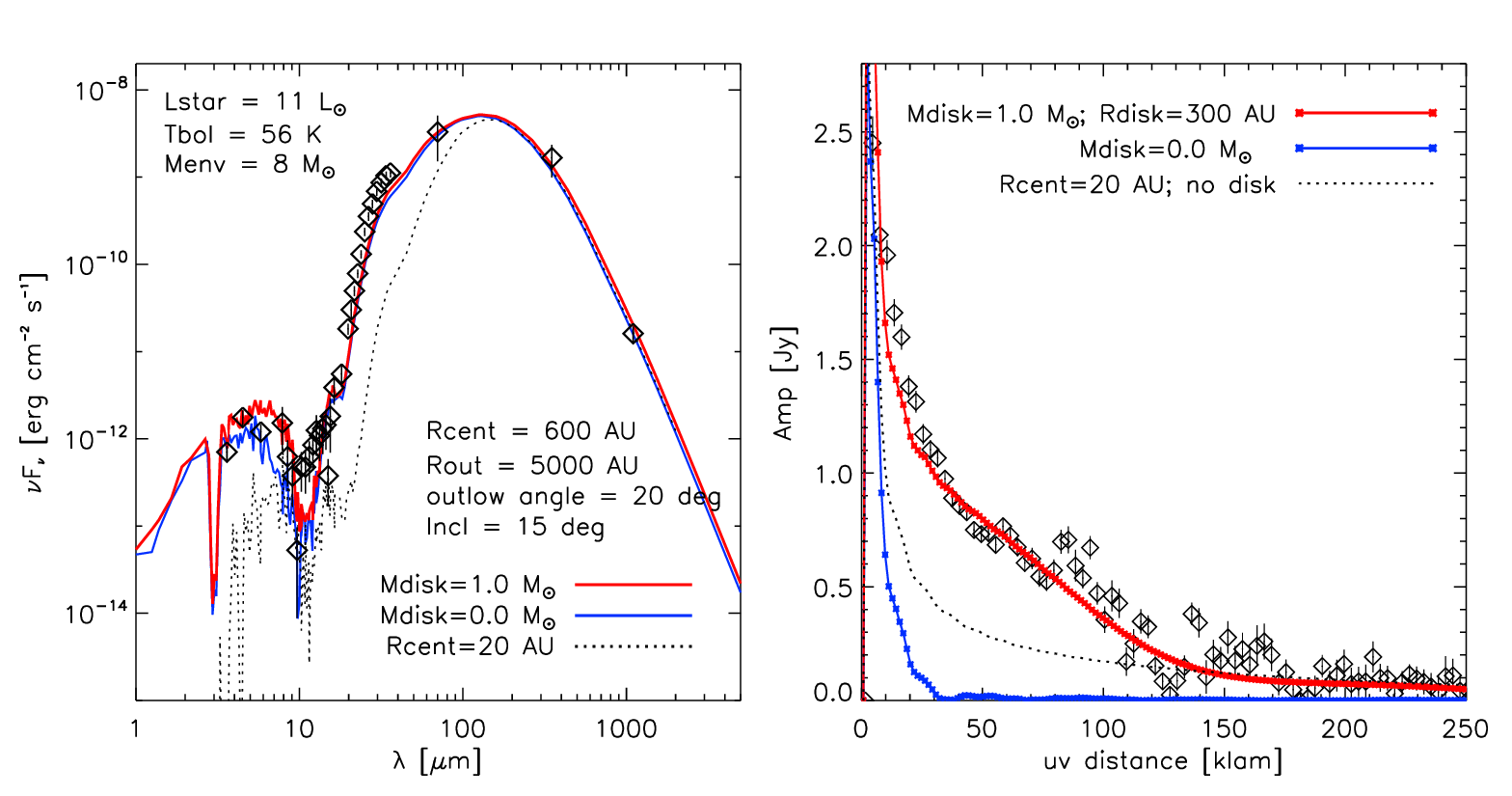}
\caption
{Best-fit envelope and disk model, compared to the observed SED (left) and 230~GHz visibilities (right).  The SED includes the binned IRS spectra and broadband data from $3.6-1100~\micron$ (Table~\ref{irstab}).  Three models are shown for comparison: the best-fit envelope and disk (red), the best fit envelope with no disk (blue), and a model with no inner envelope hole ($\rcent=20$~AU; dotted).  The envelope-only model ($\mdisk=0~\msun$) indicates the relative contributions of the envelope and disk to the 230~GHz visibility amplitudes.
\label{fitfig}}
\vspace{0.1in}
\end{figure*}

The best-fitting model ($\rout=5000$~AU, $\rcent=600$~AU, $\ang=20\deg$, $\incl=15\deg$, $\mdisk=1.0~\msun$, $\rdisk=300$~AU) is compared to the data in Figure~\ref{fitfig}.  
The same envelope model with no disk is also shown for reference, as is an envelope model with no inner envelope hole ($\rcent=20$~AU; dotted line).  
Determination of the best fit model from Figure~\ref{chifig} is described in Sections~\ref{envsec} and \ref{disksec}.  Due to the small uncertainties and the limited number of models, the reduced $\chi^2$ values for even the best-fit model are still fairly high ($\tilde{\chi}^2\sim13$ and 5 for the SED and visibilities, respectively).
The bolometric temperature and luminosity of the best-fit model are $35.3$~K and $17.3~\lsun$.  The bolometric luminosity differs from the input stellar luminosity due to inclination effects. 

While we only show the best-fit model here, there is a range of values for each parameter that can reasonably fit the data, as determined by eye from $\chi^2$ plots and visual inspection of the SED fits.  We find that the envelope parameters have the following reasonable ranges: $\rout\sim5000-7000$~AU, $\rcent\sim400-600$~AU, $\ang\sim10-30\deg$, and $\incl\sim 10-25\deg$.
Reasonable disk parameters are $\mdisk\sim 0.7-1.5~\msun$, and $\rdisk\sim 200-500$~AU.

\begin{figure}[!ht]
\hspace{-0.2in}
\includegraphics[width=3.7in]{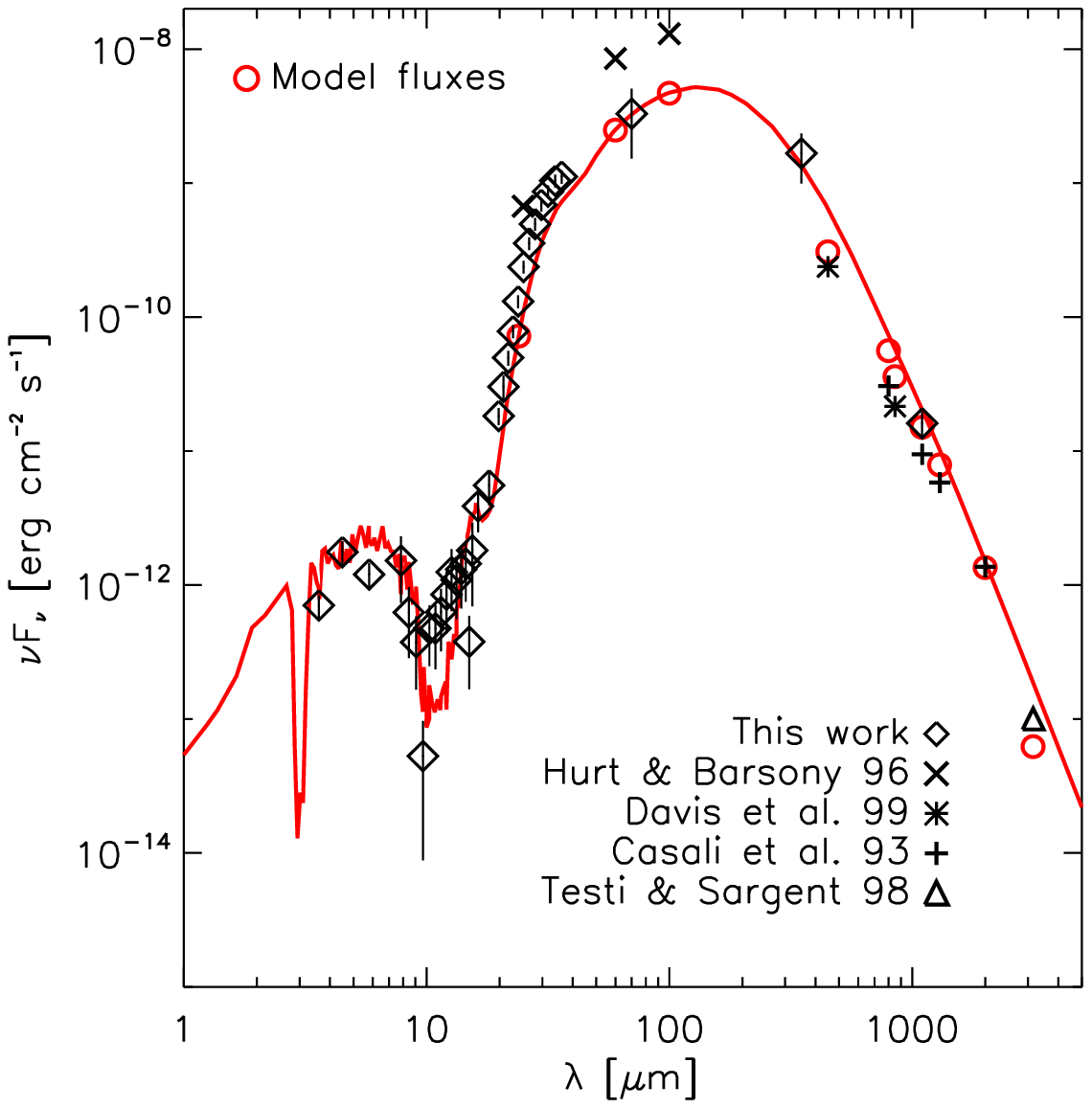}
\vspace{-0.15in}
\caption
{Best-fit model compared with literature fluxes: IRAS HIRES 25, 60, 100~$\micron$ \citep{hb96}, SCUBA 450, 850~$\micron$ (peak fluxes; \citealt{davis99}), JCMT 800, 1100, 1300, $2000~\micron$ \citep{casali93}, and OVRO 3~mm \citep{ts98}.  The model has been computed in apertures consistent with the literature measurements (circles).  IRAS fluxes are based on lower resolution data than the \textit{Spitzer} fluxes used here, and may be confused with nearby protostars.
Literature fluxes are not used in the SED fit. 
\label{litfig}}
\vspace{0.15in}
\end{figure}

Literature fluxes are shown in comparison to our data and the best-fit model SED in Figure~\ref{litfig}.  Shown are IRAS HIRES 25, 60, and 100~$\micron$ from \citet{hb96}, SCUBA 450 and 850~$\micron$ peak fluxes from \citet{davis99}, JCMT 800, 1100, 1300, and $2000~\micron$ fluxes from \citet{casali93}, and the OVRO 3~mm flux from \citet{ts98}.  They cannot be compared directly to the model SED because many of them are peak fluxes calculated in small apertures; circles show the corresponding model values computed in similar apertures.  While there is not perfect agreement, the model is roughly consistent with the literature values, with the exception of IRAS fluxes. The IRAS observations are lower resolution than the \textit{Spitzer} maps, and FIRS~1 may be confused with nearby protostars ($24-45\arcsec$ away).
Literature fluxes are not included in the $\tilde{\chi}^2$ fitting.

\subsection{Envelope Structure}\label{envsec}

Envelope parameters are determined first, using the grid of envelope models (Figure~\ref{chifig} (A)-(C)).  
The 4-D $\tilde{\chi}^2$ space is collapsed along the parameters not plotted in each panel. 
We conclude below that $\rout\le7000$~AU based on the 230~GHz visibility $\tilde{\chi}^2$ contours in panel (A), so only models with $\rout\le7000$~AU are included in panels (B) and (C) to avoid complicating the plots (because many models with $\rout>7000$~AU can fit the SED).

With a few exceptions, the SED is much more sensitive to envelope parameters than are the 230~GHz visibilities.
\rout, however, is only mildly constrained by the SED and is somewhat degenerate with \rcent. Increasing either parameter lowers the total opacity of the envelope, allowing more MIR emission to escape.  In addition, for $\rout>7000$~AU, the SED provides no constraint on \rcent\ because the opacity through the envelope is already quite low.  In this case, the 230~GHz data do help constrain the envelope parameters because visibility amplitudes at $uv$-distance $\lesssim 30 ~k\lambda$ trace extended emission.  A narrow peak in the $uv$ plane (Figure~\ref{visfig}) corresponds to a large envelope outer radius, and a wider peak to a small outer radius.  
The best compromise between the visibilities preferring smaller \rout\ and the SED preferring larger \rout\ is $\rout\ \sim 5000$~AU (Figure~\ref{chifig} A).  The value of \rcent\ with the lowest SED $\tilde{\chi}^2$ in this case is 600~AU.

A more compact envelope creates a high opacity to shorter wavelength emission, thus a larger \rcent\ is needed to decrease opacity close to the protostar and to match the observed NIR and MIR emission.   For $\rout \le7000$~AU, the range of reasonable centrifugal radii are $\rcent=400-700$~AU.  
There are a few models with small \rcent\ and low $\tilde\chi^2$; when the viewing angle is just down the edge of the outflow ($\incl=\ang/2$) in a dense envelope ($\rout=4000$~AU).  In these special cases the opacity is lowered just enough to give a similar emergent SED as models with a large inner envelope hole.

A compact envelope with $\rout\sim 5000$~AU is consistent with the crowded star formation region in which FIRS~1 is located.  The nearest embedded protostar that is known to have an envelope is $45\arcsec $, or approximately $11000$~AU away, and several other embedded sources are within a few arcminutes \citep{enoch09}.
While we do not know the actual 3-D distances, envelopes with radii $\rout\sim5000$~AU are certainly reasonable in this clustered region. 

Both the outflow opening angle and inclination are fairly narrowly constrained by the SED (Figure~\ref{chifig} (B),(C)). There is a degeneracy between \ang\ and \incl; all models where the line of sight is directly within the outflow cavity ($\incl<\ang/2$) have very high $\tilde\chi^2$ values because they produce a large excess of NIR emission.  The SED is best fit by models with low inclinations ($\incl<35\deg$); larger inclinations produce very high extinction in the MIR and cannot match the observed MIR flux.  Low inclinations may be in conflict with the orientation of the 3.6~cm jet, which has been interpreted as being almost in the plane of the sky based on proper motion of emission knots in the jet \citep{mosc06}.  

To summarize, the short $uv$-spacing visibility data favor small outer radii, while the SED favors larger outer radii, with the best compromise lying at $\rout\sim 5000$~AU.  For $\rout=5000$~AU, the SED $\tilde\chi^2$ is minimized for $\rcent= 600$~AU (panel (A) of Figure~\ref{chifig}).  With \rcent\ set to 600~AU, it is straightforward to determine the best-fit \ang\ and \incl\ from the SED $\chi^2$ curves in panels (B) and (C).

None of the models accurately reproduce the shape of the silicate absorption feature or the slope of the spectrum from $10-20\micron$, although the uncertainties on the observed fluxes are also quite high in this region.  This is most likely a feature of the dust model and not the envelope density profile.  Similarly, the dust model does not include the $15~\micron$ CO$_2$ ice absorption feature.

Note that the input stellar luminosity ($11~\lsun$) is based on the bolometric luminosity calculated before the SHARC~II $350~\micron$ was available, and thus is lower than the bolometric luminosity of the best-fitting model ($18~\lsun$).  Increasing the stellar luminosity to $18~\lsun$ does not change the results dramatically; the best-fit centrifugal radius is a bit lower, 400~AU, without changing the other parameters.  The reasonable range of $\rcent$ is much larger however, allowing for \rcent\ as low as 50~AU for larger outflow opening angles (e.g. $40\deg$).

\begin{figure*}[!ht]
\hspace{-0.2in}
\includegraphics[width=7.3in]{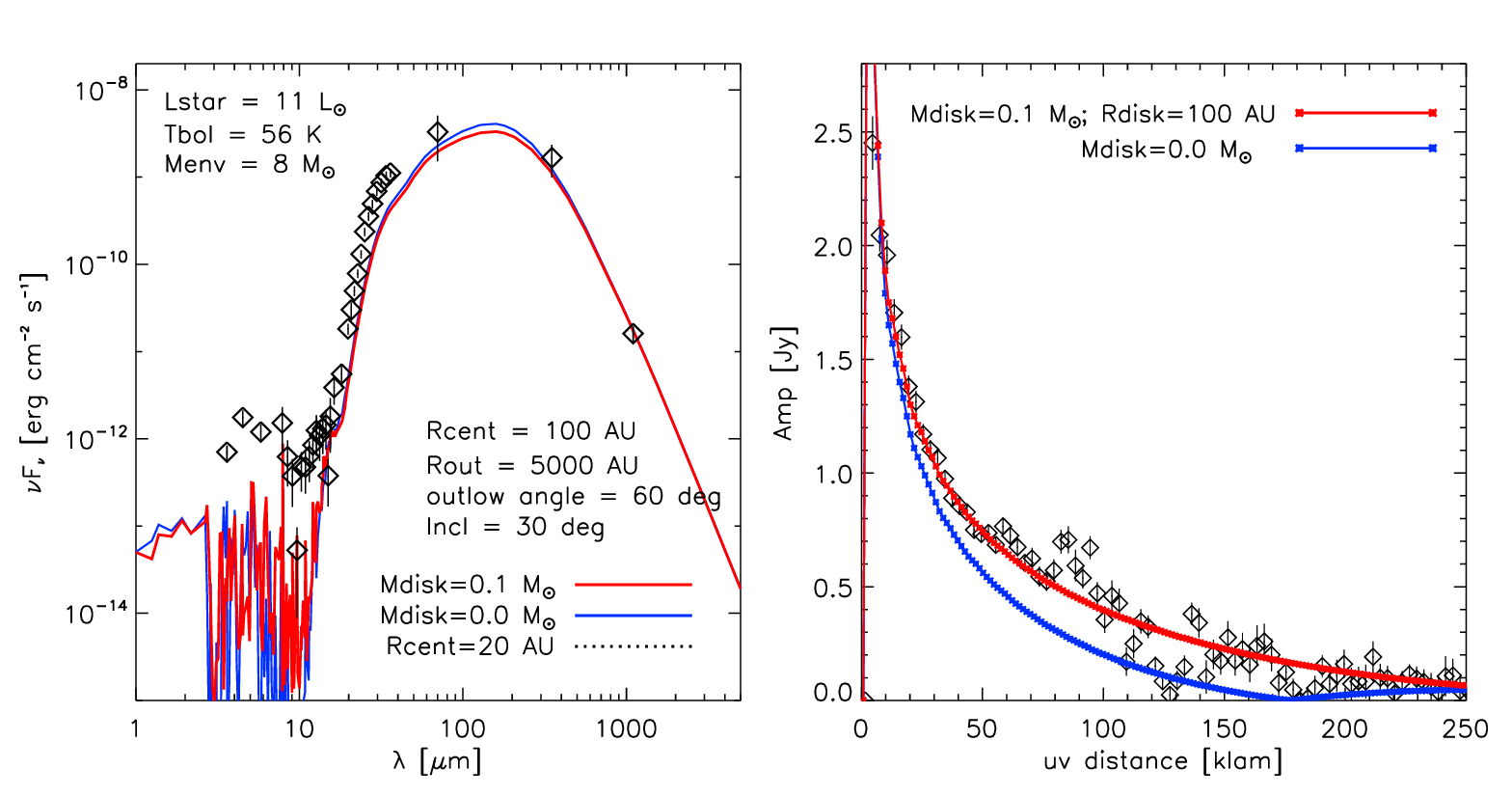}
\caption
{Best-fit power law envelope model, $\rho \propto r^{-2}$, compared to the observed SED and 230~GHz visibilities.  A steep power law alleviates the need for a massive disk ($\mdisk=0.1~\msun$, $\rdisk=100$~AU is the best fit), but the power law models are unable to match the observed MIR \textit{Spitzer} fluxes.
\label{fit2fig}}
\end{figure*}

\subsection{Disk Structure}\label{disksec}

After the envelope parameters (\rcent, \rout, \ang, \incl) have been determined, we run a separate grid in disk mass and radius with envelope parameters fixed.  The resulting $\tilde{\chi}^2$ contours are shown in Figure~\ref{chifig} (D).  Disk parameters are entirely constrained by the CARMA 230~GHz visibilities; the SED is insensitive to both \mdisk\ and \rdisk.  
A quite massive ($\mdisk \sim 0.7-1.5~ \msun$) and resolved ($\rdisk \sim 200-500$~AU, compared to the maximum resolution of 100~AU) disk is required to account for the significant flux at intermediate $uv$-distances ($20-100~k\lambda$; Figure~\ref{fitfig}).

Typically, only a lower limit can be placed on the disk mass because once the disk emission becomes optically thick larger masses do not increase the millimeter flux.  Here, however, because the disk is resolved the mass is more tightly constrained.  
For FIRS~1, fitting the visibilities out to maximum $uv$-distances from $50-300$~k$\lambda$ produces the same best-fit disk parameters, although the goodness of fit increases with more data.  The $uv$-coverage required to get a good fit should depend on the disk and envelope structure.

Disk properties derived by fitting the 230~GHz visibilities are relatively insensitive to the assumed envelope parameters.  It is true that a somewhat less massive disk would be required if there were no inner envelope hole; for example for $\rcent=50$~AU, the best-fit is for $\mdisk\sim0.4$, $\rdisk\sim400$~AU.
However, for envelope parameters with any reasonable fit to the SED (for example, $\rcent=400$~AU, $\rout=7000$~AU), $\mdisk=1.0~\msun$ and $\rdisk=300$~AU remain the best fit to the 230~GHz visibilities.

The disk mass is also reasonably robust to uncertainties in the envelope mass ($8~\msun$) and 230~GHz calibration.  For $\menv=10~\msun$ the best-fit disk is unchanged, because only the amplitude of the narrow peak at small uv-distances 
is affected by the envelope mass.  For $\menv=6~\msun$, a slightly larger, more massive disk ($\mdisk\sim1.5~\msun$, $\rdisk\sim400$~AU) is required to account for the decrease in envelope flux at small uv distances.  The overall fit is poorer than for $\menv=8~\msun$, however.
Systematic uncertainties in the CARMA 230~GHz fluxes have a slightly larger effect, with a 30\% change in overall flux calibration producing a corresponding 30\% change in the disk mass: $\mdisk\sim 0.5-0.7~\msun$ for a 30\% decrease, and $\mdisk\sim 1.5-2.0~\msun$ for a 30\% increase.

\subsection{Other models}\label{robsec}

Here we compare our models to other disk and envelope models.  This serves both as a check on our derived envelope and disk parameters by indicating which conclusions are dependent on the density model, and a test that our models give the most reasonable fit to the data.

\citet{hoger99} observed FIRS~1 with the OVRO interferometer at 3.4, 2.7, and 1.4 mm.  They used a power law envelope model plus a point source, with the dust temperature power law set at $-0.4$ and fixed inner and outer envelope radii of 100 and 8000 AU.  These millimeter data were best fit by an envelope with mass $6 ~\msun$ and density power law $-2.0$, plus an unresolved point source of approximately $0.7~\msun$.
\citet{hoger99} note that they are unable to separate the inner envelope from any disk emission, and thus cannot place a meaningful limit on the disk itself.  Compared to the \citet{hoger99} $uv$-coverage, $10-180~ k\lambda$ at 1.4 mm, our CARMA observations trace much more of the envelope (down to $4.5 ~k\lambda$), allowing us to separately model and remove the envelope contribution.  Our data also have much higher signal-to-noise on long baselines ($uv$-distances $> 100 ~k\lambda$).  

In addition to the rotating, collapsing spheroid (or ``Ulrich'') envelope models described in Section~\ref{modsec}, we also ran a small grid of models with a simple power law envelope density profile ($\rho \propto r^{-p}$), plus a conical cavity.  A steep power law, $\rho \propto r^{-2}$, provides a reasonable fit to the visibilities without requiring a massive disk or large inner envelope hole.  
The best fit is for $\mdisk=0.1~\msun$, $\rdisk=100$~AU, and $\rcent=100$~AU, as shown in Figure~\ref{fit2fig}.  Here the emission at intermediate $uv$-distances is filled in by the envelope, which reaches high densities close to the protostar, thus requiring less disk emission. 
But, only a special combination of outflow opening angle and
inclination can come close to matching the SED ($\ang=40\deg$,
$\incl=25\deg$ or $\ang=60\deg$, $\incl=30\deg$; looking down the edge
of a wide outflow cavity), and even the best-fit model gives a much
poorer fit to the SED than the Ulrich models.  In general, the power
law models seem unable to reasonably fit the NIR and MIR emission,
although only $p=-2$ and $-1.5$ have been tested here.

The ability of the power law envelope model to fit the visibilities without a large disk is consistent with some previous studies that have found that disks are often not required to millimeter data of Class 0 and Class I sources \citep[e.g][]{loon00}.  The results here demonstrate that it is necessary to include both spectral and visibility data in order to fit a consistent disk and envelope model.

We use the online SED fitting tool of \citet{rob07} as another estimate of the envelope parameters, although the 230~GHz visibilities cannot be included in the fit.  \citet{rob07} use an envelope setup similar to ours, with the envelope infall rate $\dot{M}_{env}$ setting the fiducial density $\rho_0$ (rather than $\menv$ as used here).  
The best-fit model corresponds to a protostar with age $t=2\times10^5$~yr, $M_*=1.8 ~\msun$, $R_{*}=7~\rsun$, $T_*=4400$~K, envelope infall rate $\dot{M}_{env}=10^{-4} ~\msun$~yr$^{-1}$, $\rout=11000$~AU, $\ang=27\deg$, and $\incl=75\deg$.
The total luminosity and envelope mass of this best fit model are
consistent with our values (18.4~\lsun\ and 7.4~\msun).  

Looking at the 10 best fitting models, only the age ($t<2\times10^5$~yr), protostellar mass ($M_*<2~\msun$), temperature ($T_*=3000-4500$~K), and envelope infall rate ($\dot{M}_{env}=10^{-5}-10^{-4}~\msun$~yr$^{-1}$) are reasonably well constrained, while the other parameters cover a large range.
We do not attempt to constrain the disk properties as the SED is relatively insensitive to the disk in embedded sources.  In addition, we cannot use the online grid to constrain the envelope inner radius, because it is fixed to the disk inner radius and very few models with both large envelope mass ($>1~ \msun$) and large inner envelope radius ($\gtrsim 10$~AU) are included in the grid.  
This difference in the inner envelope behavior likely accounts for the large outer radius and inclination required by the \citet{rob07} models.

Given our limited exploration of various models, we feel that the Ulrich envelope model provides the best description of the observed SED.  While the derived disk parameters do depend on the input envelope density profile, even in the most conservative case $\mdisk\gtrsim0.1~\msun$.

\section{Discussion}\label{discsec}

Our derived disk mass of $\mdisk \sim 1.0~ \msun$ within a radius of 300~AU is consistent with the \citet{brown00} limit of $\mdisk>0.1~\msun$, as well as the \citet{hoger99} limit of $0.7~\msun$ on the unresolved mass within a radius of 100~AU.  
The early evolutionary state of FIRS~1 is confirmed by the low bolometric temperature ($\tbol\sim 56$~K) and the small disk-to-envelope mass ratio ($\mdisk/\menv \sim 0.1$) despite the high disk mass.  Thus, our results suggest that large disks can accumulate very early in the protostellar collapse process.
  
The FIRS~1 disk is likely too small to be considered a magnetically supported ``pseudo-disk''.
Given the expected young age of FIRS~1, however, both the mass and radius derived here are much larger than expected for disk formation via gravitational collapse of a rotating core.  
\citet{tsc84} predict that the disk radius, where centrifugal balance is achieved, should depend on the initial rotation $\Omega$ and isothermal sound speed $c_s$ in the core as:
\begin{equation}
R_d = 7\left(\frac{c_s}{0.35 \mathrm{km~ s}^{-1}}\right) \left(\frac{\Omega}{4\times10^{-14} \mathrm{s}^{-1}}\right)^2 \left(\frac{t}{10^5 \mathrm{yr}}\right)^3 \mathrm{AU}. 
\end{equation}

Based on the statistical relationship between \tbol\ and time derived in \citet{enoch09} ($\tbol \propto t^{1.8}$), the bolometric temperature of FIRS~1 suggests that it has an age of $0.7-0.8 \times 10^5$~yr.
For a reasonable sound speed, $c_s\sim0.23$~km~s$^{-1}$ ($T\sim15$~K), this age and
$\rdisk=300$~AU requires an initial rotation rate of approximately
$5\times10^{-13}$~s$^{-1}$.  This value is higher than typical dense cores,
which have $\Omega\sim 10^{-13} - 10^{-14}$~s$^{-1}$
\citep{goodman93}.  
Alternatively, if the age of the source is
actually closer to $3\times10^5$~yr, a more typical rotation rate
would be sufficient for growing a 300~AU disk.

Similarly, we can estimate how long it would take for the disk to build up $1~\msun$ via infall from the envelope.  For an infall rate of $\dot{M}_{env} \sim c_s^3/G \sim 10^{-5}~\msun$~yr$^{-1}$, and conservatively assuming that all of the infalling material falls onto the disk rather than directly onto the protostar, accumulating $1~\msun$ would take $10^5$~yr.  Although this is probably close to the age of FIRS~1, a disk of $1~\msun$ requires that little of the infalling material be accreted from the disk onto the star.
Below we mention a few plausible methods for building up a large circum-protostellar disk in this object.

A Class~0 lifetime longer than a few times $10^5$~yr, i.e., longer than the estimated timescale for Class 0 sources in nearby low mass star forming regions \citep{enoch09,evans09}, would allow larger disks to grow before the end of the Class~0 phase.

FIRS~1 may have a higher envelope infall rate than average, allowing the disk to quickly accumulate mass.  The bolometric luminosity of FIRS~1 is quite large compared to the typical luminosity of YSOs in nearby low-mass star forming regions ($\lesssim1\lsun$; \citealt{dun08,enoch09}).
A luminosity of $11~\lsun$ implies an accretion rate onto the protostar of at least $2\times10^{-5}~\msun$~yr$^{-1}$ (for $\dot{M}_{*} \sim R_*L_{bol}/GM_*$, $R_*\sim 5~R_{\odot}$ and $M_*\sim1~\msun$).  If this corresponds to an even higher envelope infall rate, $\mdisk \sim1~\msun$ could easily be achieved in $0.7-0.8\times10^5$~yr.

If the FIRS~1 disk has very low viscosity, mass may build up in the disk with very little accreting onto the protostar (although this is at odds with the high luminosity).  \citet{brown00} suggest that early disk formation and similar disk masses in the Class~0 and Class~I phases could be achieved with a time-dependent viscosity, low at early times and higher by Class~I.

Perhaps more likely, this source could have recently entered a period of relatively rapid accretion, as expected in the episodic accretion scenario \citep[e.g][]{hk85,enoch09,evans09}.  Such a high mask disk around a presumably low protostar mass should be unstable, and undergoing rapid accretion, explaining the high luminosity.  In this picture, the current high accretion phase would have been preceded by a period of low accretion onto the protostar while material built up in the disk (assuming infall from the envelope onto the disk is steady).

A larger sample is certainly needed to determine if such large, high mass disks are typical in the Class~0 phase.
The recent \citet{jorg09} study of 20 Class~0 and Class~I sources finds disks masses in Class~0 from $0.01-0.5~\msun$, and $\mdisk/\menv$ ratios of $1-10$\%.  If we calculate our disk mass by the same method as \citet{jorg09}, which uses the flux at $50$~k$\lambda$ and assumes an optically thin, unresolved disk, we get $\mdisk=0.6~\msun$.  
While this is at the high end of the \citet{jorg09} disk sample, the disk to envelope mass ratio ($\sim8$\% when using $\mdisk=0.6~\msun$) is consistent with their results, as is the idea that disks are already well established in the Class~0 stage. 

Regarding the envelope structure, it is important to keep in mind that while we refer to \rcent\ as the centrifugal radius, our model is not dynamical, and there is no dependence on rotation rate.  Thus the sharp drop in density inside of this radius could have any number of causes, including a companion that has cleared out material, as well as rotational collapse onto a disk. 
Any binary companion with a disk mass larger than 0.1~\msun\ should have been detected.  There is a tentative second detection 500~AU to the northwest of FIRS~1 (see Figure~\ref{carmafig} C); the peak of approximately 55~mJy would correspond to a disk mass of $\sim0.06~\msun$, but this may just be a ``clumpy'' feature in the envelope.
The more likely explanation is that the inner envelope cavity is a result of collapse in a rotating core, and creation of the 300~AU radius disk.
Alternatively, a clumpy envelope could cause a similar decrease in MIR opacity and might alleviate the need for an inner envelope hole \citep{indeb06}.

If the disk and envelope are physically connected, with both the disk and inner envelope hole governed by rotation in the collapsing core, we might expect $\rdisk \approx \rcent$.  Although the best-fit \rcent\ is a factor of two larger than \rdisk\ here, the range of reasonable values allow for \rcent\ to be as small as 400~AU and \rdisk\ to be as large as 500~AU.  Thus, a physical continuity between the disk and envelope is certainly plausible.

\section{Summary}\label{concsec}

We utilize \textit{Spitzer} IRS spectra, high resolution CARMA 230~GHz
continuum data, and broadband photometry together with a grid of
radiative transfer models to characterize the disk and envelope
structure of the Class~0 protostar Serpens FIRS~1.  
Our conclusions are:

1. Radiative transfer models combined with mid-infrared spectra and millimeter data with excellent $uv$-coverage can reasonably constrain envelope parameters, including the inner (centrifugal) radius, outer radius, and outflow opening angle.  In all cases there is a range of parameter values able to reasonably fit the data.  
Once the envelope parameters have been determined, the mass and radius of the disk are robustly constrained by millimeter interferometry data with $uv$-coverage from $<5$ to $>300~k\lambda$.

2. We find a centrifugal radius for FIRS~1 in the range $400-700$~AU,
indicating a large ``hole'' in the inner envelope, similar to IRAS~16293 \citep{jorg05}. Unlike IRAS~16293, however, there is no strong evidence for a binary companion that might have cleared out the inner envelope.  Other explanations for such a large \rcent\ in this source are: (a) collapse of the inner envelope onto the disk due to the conservation of angular momentum in a rotating, collapsing core, or (b) the \rcent\ does not indicate a true inner radius, but rather a ``clumpy'' envelope with much lower opacity in the MIR than a smooth envelope density profile.

3. Using envelope parameters set by the SED, the CARMA 230~GHz visibilities require a quite massive, resolved disk.  The best-fitting disk has a mass of $1\msun$, and a radius of approximately $300$~AU.
While this mass is consistent with previous limits \citep{hoger99,brown00}, it also indicates that protostars can accumulate relatively massive disks at very early times.  This is somewhat at odds with theoretical expectations that disks start small and grow with time \citep{tsc84}.
The range of reasonable disk and envelope parameters does allow for a physical continuity between the disk and inner envelope.

4. Our results for FIRS~1 demonstrate the feasibility of using this method to characterize the disk and envelope structure in a larger sample of Class~0 sources.  Similar modeling for an unbiased sample will allow us to characterize the typical disk mass, size, and inner envelope structure during the Class 0 phase.

\acknowledgments

The authors thank the referee T. Bourke, and N. Evans for comments and suggestions that improved this manuscript.  We are also grateful to C. Dullemond for the use of his RADMC radiative transfer code and helpful discussions, and to A. Crapsi for providing his model setup.
Support for this work was provided by NASA through the Spitzer Space Telescope Fellowship Program, through a contract issued by the Jet Propulsion Laboratory (JPL), California Institute of Technology, under a contract with NASA.
Support for CARMA construction was derived from the states of California, Illinois, and Maryland, the Gordon and Betty Moore Foundation, the Kenneth T. and Eileen L. Norris Foundation, the Associates of the California Institute of Technology, and the National Science Foundation. Ongoing CARMA development and operations are supported by the National Science Foundation under a cooperative agreement, and by the CARMA partner universities.
Support for c2d, a Spitzer Legacy Science Program, was provided by NASA through contracts 1224608 and 1230782 issued by JPL, California Institute of Technology, under NASA contract 1407.
The development of Bolocam was provided by NSF grants AST-9980846 and AST-0206158.

\end{document}